\newcommand{\ket}[1]{|{#1}\rangle}
\newcommand{\bra}[1]{\langle{#1}|}

\documentclass[epj]{svjour}

\usepackage{epsfig}                                                            
\usepackage{amssymb}                                                           
\usepackage{amsmath}                                                           
\usepackage{graphicx}                                                          

\begin{document}

\title{A simple quantum gate with atom chips}
\author{M. A. Cirone\inst{1,2} \and A. Negretti \inst{1,2,3} \and T. Calarco \inst{1,2} \and P. Kr\"uger \inst{4} \and J. Schmiedmayer \inst{4}}

\institute{Dipartimento di Fisica, Universit\`a di Trento, and CRS
INFM-BEC, I-38050 Povo, Italy \and ECT*, Strada delle Tabarelle
286, I-38050 Villazzano, Italy \and Institut f\"ur Physik,
Universit\"at Potsdam, Am Neuen Palais 10, D-14469 Potsdam,
Germany, \and Physikalisches Institut, Universit\"at Heidelberg,
Philosophenweg 12, D-69120 Heidelberg, Germany}

\abstract{We present a simple scheme for implementing an atomic
phase gate using two degrees of freedom for each atom and discuss
its realization with cold rubidium atoms on atom chips. We
investigate the performance of this collisional phase gate and
show that gate operations with high fidelity can be realized in
magnetic traps that are currently available on atom chips.
} \maketitle

\section{Introduction}

Neutral atoms are promising candidates for the physical
implementation of quantum information processing (QIP). The weak
interaction of neutral atoms with their environment leads to long
coherence times, and mature experimental techniques from quantum
optics and atomic physics allow to prepare and manipulate atomic
quantum systems.

In recent years new tools for the precise control and manipulation
of neutral atoms were developed, based on the adaption of
microfabrication techniques to atom optics and the implementation
of the {\em atom chip} \cite{Fol02}. Such atom chips apply the
advantages of micro-fabrication technology from micro-electronics
and micro-optics to atomic phys\-ics with the goal to build
integrated devices for quantum manipulation of ultra-cold atomic
samples and a collection of single atoms.

In the quest for implementations of QIP with neutral atoms, atom
chips are promising candidates for a number of reasons: (i) Large
electric and magnetic field gradients and field curvatures near
microscopic conductors lead to tight confinement and large energy
level spacings for the trapped atoms (trap frequencies in the
range of hundreds of kHz were implemented \cite{Fol00}). (ii) Very
high resolution of the potentials (sub-$\mu$m) can be implemented
when going close to the surface ($\mu$m distance). This is
important for QIP proposals in which the short distances (order of
1 $\mu$m) between individual trapping sites is required to achieve
sizable qubit coupling and fast gate operations. (iii)
Nanofabrication techniques on surfaces facilitate accurate and
robust placement of structures with size limit well below 100 nm.
This enables the realization of nearly arbitrary potential
configurations. (iv) Integration of new components, for example
micro-optics and micro-cavities for preparation, manipulation, and
detection of qubits are possible. (v) Last, but not least,
nanofabrication schemes are particularly well suited for
production of multiple structures.

In this paper we discuss in more detail how to implement a
quantum logic operation with the toolbox of atom chips. We will
make thereby use of two different degrees of freedom, each of
them having two levels, for each qubit. We use the first degree
of freedom to storage information and the second one to process it.
This helps to take advantage of the best features of both degrees
of freedom. One pair of states (the hyperfine levels of a trapped
atom) will be used to storage the qubit, and the quantum gate
operations will make use of two vibrational levels of the atom
trap as qubit states and the collisions that realize the phase
gate will occur in the same internal states, which has the
advantage of reducing unwanted collisional losses.  Our scheme
therefore combines the important feature of long lived coherence
of the hyperfine states with the phase gate operation with
vibrational states via collisions.

We shall first discuss the implementations of qubits on atom
chips, and then describe new schemes for implementing a phase gate
for cold neutral atoms on atom chips. The results of a realistic
calculation of the gate performance using typical state of the art
atom chip parameters is given in Section \ref{magn}. Section
\ref{futu} briefly illustrates future directions of QIP on atom
chips.

\section{The Qubit}
\label{qub} We shall discuss here only implementations where each
qubit is written into a single atom. There are two distinct ways
to encode a qubit into a single neutral (Rb) atom on atom chips:

\begin{itemize}
\item The qubit is encoded into a pair of long-lived internal states
such as two different hyperfine ground states of an atom.  The qubit
states have to be trappable on the atom chip, and show long coherence
times.

To achieve long coherence times, it is advisable to use atomic
states where the energy difference is to first order independent
of external fields (clock states). In free space these would be
the $\ket{F=2, m_F=0}$ and $\ket{F=1, m_F=0}$ states used in
atomic clocks, which show only quadratic Zeeman and Stark shifts.
On the atom chip one can achieve the same common mode rejection of
external field noise using the hyperfine levels $\ket{F=2, m_F=1}$
and $\ket{F=1, m_F=-1}$ of the $5S_{1/2}$ ground state of
$^{87}$Rb which have the same magnetic moment at a magic magnetic
field of $B_M = 3.23$ G \cite{Cornell}. Operating a trap at this
offset field, the energy difference between these two trapable
states shows only a quadratic dependence on external field
fluctuations like the $m_F = 0$ clock states at zero field.

Single-qubit operations are induced as transitions between the
hyperfine states of the atoms. These can be driven by external
fields, using radiofrequency (RF) and/or microwave (MW) pulses or
optical Raman transitions. As an example we mention the atom chip
experiment in Munich \cite{rei1} which demonstrated the very long
coherence times in a magnetic atom chip trap for the special clock
states $\ket{F=2, m_F=1}$ and $\ket{F=1, m_F=-1}$ at a trapping
magnetic field near the value $B_M$.

\item The qubit is encoded in external, motional states of the atom in a tight
trap. These motional states can be either the ground and an
excited state in a trap, or the left and right states of a double
well. In such a realization the atoms are in the same
internal state, and therefore automatically isolated from external
fluctuations. Single-qubit operations are then Rabi rotations between
the motional states, and can be seen like trapped atom interferometers.
\end{itemize}

To implement a quantum bit on an atom chip, the qubit states have
to be trappable, and show long coherence times. That is we require
a common mode rejection of external field noise (clock states).
Atoms trapped in different motional states are easier to
state-selectively manipulate, but much less is known on external
state decoherence, and how to prevent it.

\section{New schemes for a phase gate}
\label{sche}

The phase gate is a two-qubit quantum gate described by the evolution operator \cite{nielsen}

\begin{eqnarray*}
\label{matrixphasegate}
\mathcal{H}
(\phi)=\left(
\begin{array}{cccc}
1 & 0 & 0 & 0 \\
0 & 1 & 0 & 0 \\
0 & 0 & 1 & 0 \\
0 & 0 & 0 & e^{i\phi}
\end{array}
\right)
\end{eqnarray*}

Several different implementations of this universal gate have been
proposed. In particular, schemes where internal electronic states
of atoms are the logic states $\ket{0}$ and $\ket{1}$ and
state-depend\-ent atomic collisions generate appropriate values of
the phase $\phi$ (usually $\pi$) have been discussed
\cite{cala,tren,delo,will}. In these cases the correct performance
of the phase gate requires that at some time $\tau$ all
vibrational states have a complete revival up to a phase.

Our schemes use two different degrees of freedom, each of them
having two levels, for each qubit. We use the first degree of
freedom to storage information and the second one to process it. The
``storage'' qubit levels are denoted as $\ket{0}$ and $\ket{1}$,
whereas the ``operation'' qubit levels are denoted as $\ket{g}$
and $\ket{e}$. Such a scheme helps to take advantage of the best
features of both degrees of freedom.

In principle, the qubit operations we shall describe can be
performed directly on the storage states, so our schemes might
appear as an unnecessary complication.

In the specific scheme presented here, the storage qubit is encoded
in the clock states $\ket{0} \equiv \ket{F=2, m_F=1}$ and $\ket{1}\equiv \ket{F=1, m_F=-1}$ of
the $5S_{1/2}$ ground state of $^{87}$Rb. Quantum coherences
between these two states have been demonstrated with decoherence
times exceeding 1 s \cite{rei1} when the value of the trapping
magnetic field is set to the magic value $B_M$. However, these
clock qubit states are trapped in identical trapping potentials.
Direct realizations of the phase gate via internal state-dependent
time-varying potentials like the one described in \cite{cala} are
hence not feasible. For this reason, we choose two vibrational
states as operation qubit states as suggested in \cite{delo,will}
in order to realize a collisional phase gate that does not rely on
the internal states.

This approach combines the important feature of long coherence
time of the hyperfine states with the phase gate operation with
vibrational states via collisions. The storage states are the
hyperfine levels $\ket{F=2, m_F=1}$ and
$\ket{F=1, m_F=-1}$, the operation states $\ket{g}$
and $\ket{e}$ are the ground and first excited states of the
atomic vibrations, respectively. The MW-RF two-photon transitions
described in \cite{rei1} can be used for single qubit operations
on the storage states. For a two-qubit phase gate, our schemes can
be summarized as follows: the logic state is encoded and stored in
the hyperfine states; when two-qubit gate operation must be
performed, the logic state is transferred into the vibrational
states and the gate operation takes place via collisions; at the
end of the operation, the logic state of the vibrations is
transferred back into the internal hyperfine states.

There are two different ways to realize this idea. We can (a)
duplicate the logic state of the storage levels in the vibrational
levels or (b) swap the logic states of the two degrees of freedom.

The duplication scheme (a) for the phase gate is summarized by the
map

\begin{eqnarray}
| 0g \rangle & \rightarrow & | 0g \rangle \\
| 0e \rangle & \rightarrow & | 0e \rangle \\
| 1g \rangle & \rightarrow & | 1e \rangle \\
| 1e \rangle & \rightarrow & | 1g \rangle
\end{eqnarray}
whereas the swap scheme (b) is summarized by

\begin{eqnarray}
| 0g \rangle & \rightarrow & | 0g \rangle \\
| 0e \rangle & \rightarrow & | 0e \rangle \\
| 0e \rangle & \rightarrow & | 1g \rangle \\
| 1g \rangle & \rightarrow & | 0e \rangle
\end{eqnarray}

Let's assume that information is initially encoded in two storage
levels of two qubits,

\begin{equation}
\ket{\varphi_0} = \left( a\ket{00} + b\ket{01}
+ c\ket{10} + d\ket{11} \right) \otimes \ket{gg}
\label{init}
\end{equation}
which may be already entangled. The duplication scheme
takes place in three steps: (i) We selectively excite the (vibrational) operation state,

\begin{equation}
\ket{\varphi_1} = a\ket{0g, 0g} + b\ket{0g, 1e}
+ c\ket{1e, 0g} + d\ket{1e, 1e}
\end{equation}
(ii) only the two $\ket{e}$ states collide, acquiring a dynamical phase equal to $\pi$

\begin{equation}
\ket{\varphi_2} = a\ket{0g, 0g} + b\ket{0g, 1e} + c\ket{1e, 0g} -
d\ket{1e, 1e} \label{Eq:store}
\end{equation}
like in \cite{cala,tren,delo,will}, where only the two excited
(vibrational or internal) states get a collisional phase on top of the kinematic phase;
finally (iii) we de-excite (selectively) the operation state:

\begin{equation}
\ket{\varphi_3} = \left( a\ket{00} + b\ket{01}
+ c\ket{10} - d\ket{11} \right) \otimes \ket{gg}
\end{equation}
In this way the result of the phase gate operation on the operation states is transfered
to the storage states. The second alternative scheme swaps the storage and operation states
for the phase gate performance. Starting from the initial state Eq.(\ref{init}), again three
steps are required for the gate operation: (i) we selectively excite the operation state
and de-excite the storage states

\begin{equation}
\ket{\varphi_1'} =\ket{00} \otimes \left( a\ket{gg} + b\ket{ge}
+ c\ket{eg} + d\ket{ee} \right)
\end{equation}
i.e, we swap their logic states; then (ii) the operation states get a dynamical phase equal to $\pi$

\begin{equation}
\ket{\varphi_2'} =\ket{00} \otimes \left( a\ket{gg} + b\ket{ge}
+ c\ket{eg} - d\ket{ee} \right)
\end{equation}
through collisions as in the duplication scheme; finally (iii) we swap again the storage and operation states

\begin{equation}
\ket{\varphi_3'} =\left( a\ket{00} + b\ket{01}
+ c\ket{10} - d\ket{11} \right) \otimes \ket{gg}
\end{equation}
These schemes are not restricted to internal and external degrees of freedom
of cold atoms, but can be applied to any system with at least two degrees of freedom.

The duplication scheme does not modify the storage states. During
the phase gate operation (Eq.\ \ref{Eq:store}), the storage and the
operation states are entangled. In the swap scheme, on the other
hand, the storage states are modified but the two (storage and
operation) degrees of freedom remain always separable.

\section{Excitations of vibrations of hyperfine states of neutral atoms}
\label{exci}

The two schemes for a phase gate described in the previous section
require a selective excitation of vibrational states when
implemented with cold atoms. We specialize our discussion to
$^{87}$Rb atoms, having mass $M=1.44 \times 10^{-25}$ kg, confined
in traps of frequency $\nu_{\rm t}=10$ kHz and higher. Two-photon
Raman processes have already been used for sideband excitations of
trapped ions in the Lamb-Dicke regime \cite{wine} and are natural
candidates also to excite vibrational states of neutral trapped
atoms. So we examine the transitions from an initial state $\mid i
\rangle$ to a final state $\mid f \rangle$ driven by two external
radiation fields through an intermediate state $\mid b \rangle$
which is never populated. From the Hamiltonian of a three-level
atom interacting with two radiation fields an effective two-level
Hamiltonian

\begin{equation}
H_{\rm eff}=\hbar \omega_i \ket{i}\bra{i}+\hbar \omega_f \ket{f}\bra{f} +
\frac{\hbar \Omega_0}{2} \left[ e^{ikx}\ket{f}\bra{i} + \mbox{h.c.} \right]
\label{ef}
\end{equation}
is obtained, where $k_1-k_2 < k < k_1+k_2$ and $k_1$ and $k_2$ are the wave vectors of the two driving fields,
$\Omega_0=\Omega_1 \Omega_2/(2 \Delta)$ is the effective Rabi frequency,
$\Delta$ is the detuning of the two fields from the transition frequencies
$\omega_{bi}$, $\omega_{bf}$. The factors $\Omega_{1},\Omega_{2}$ are the Rabi frequencies
of the transitions $\ket{i} \leftrightarrow \ket{b}$ and $\ket{f} \leftrightarrow \ket{b}$.
The effective Hamiltonian (\ref{ef}) is obtained under the hypotheses
$\Omega_1, \Omega_2 \ll \Delta$.
The atomic motion is quantized when we write

\begin{equation}
kx=\eta (a^\dagger + a)
\end{equation}
where $a$ and $a^\dagger$ are the usual annihilation and creation operators of the
harmonic oscillator and

\begin{equation}
\eta\equiv \frac{\hbar k}{\sqrt{2 M\hbar\omega_{\rm t}}}
\end{equation}
is the Lamb-Dicke parameter.

The two-photon process can be obtained either through MW and RF
transitions involving other hyperfine levels of the $5S$ ground
state or through optical transitions involving a hyperfine level
of the $5P$ excited state. Microwave transitions between the $5S$
hyperfine states occur at the frequency $\nu_{\rm hf}\simeq 6.835$
GHz, so $\eta_{MW}$ is of the order of $10^{-5}$, well within the
Lamb-Dicke limit. The effective Hamiltonian Eq.(\ref{ef}) can then
be approximated by a Jaynes-Cummings Hamiltonian
\begin{eqnarray}
H_{\rm eff,JC} & = & \sum_n \hbar \omega_n \ket{\varphi_n}\bra{\varphi_n}
+\hbar \omega_i \ket{i}\bra{i}+\hbar \omega_f \ket{f}\bra{f} \nonumber \\
& & +\frac{\hbar \Omega}{2} \left[ a^\dagger \sum_{j}\ket{f,\varphi_j} \bra{i,\varphi_j} + \mbox{h.c.} \right]
\label{effeJC}
\end{eqnarray}
where the rotating wave approximation has been used and the
eigenstates $\ket{\varphi_j}$, with energies $\hbar \omega_j$, of
the atomic motion have been included \cite{comm}. In
Eq.(\ref{effeJC}) the sideband Rabi frequency $\Omega=\Omega_0
\eta_{MW} \ll \Omega_0$ replaces the effective Rabi frequency
$\Omega_0$. Since the effective Rabi frequency $\Omega_0$ can
easily be increased up to $\Omega_0 \simeq 2 \pi \times 100$ kHz
\cite{treu}, the sideband Rabi frequency has an upper bound of the
order of $2 \pi $ Hz. Microwave transitions would then occur
rather slowly.

Optical transitions employing the $5P_{1/2}$ or $5P_{3/2}$ level
as intermediate state have wavelengths of the order of
$\lambda_{\rm opt} \simeq 800$ nm. The upper bound of the
Lamb-Dicke parameter is $\eta_{\rm opt}\simeq 1$. The Lamb-Dicke
regime necessary for sideband excitation requires either the
choice of appropriate directions of the the wave vectors ${\bf
k}_1$ and ${\bf k}_2$ of the two radiation fields or trapping
frequencies above $10^5$ Hz. The ongoing miniaturizations of atom
chip structures have reached this range of frequencies
\cite{Fol00}.

In the duplication scheme only the vibrational state must be
modified. The desired transition is $\ket{1g} \leftrightarrow
\ket{1e}$, the energy difference of which gives the sideband
excitation condition $k_1-k_2=2.09 \times 10^{-6} {\rm cm}^{-1}$.
Microwave Raman processes where the $5S(F=2,m_F=-2)$ ground state
is the intermediate state would ensure selectivity, but the Rabi
oscillations would be slow, as mentioned before. On the other
hand, the optical transitions do not fulfill the selectivity
requirement, since the lasers that drive the transitions $\ket{1g}
\leftrightarrow \ket{1e}$ would also drive the undesired
transitions $\ket{0g} \leftrightarrow \ket{0e}$ (only the detuning
$\Delta$ of the two transitions would be slightly different). The
duplication schemes presents severe drawbacks for the system we
are considering.

In the swap scheme both the internal and the vibrational states
are changed, so the desired sideband excitation $\ket{1g}
\leftrightarrow \ket{0e}$ requires $k_1-k_2 = 2 \pi (\nu_{\rm
t}+\nu_{\rm hf})/c \simeq 2 \pi \nu_{\rm hf}/c=1.43 \; {\rm
cm}^{-1}$. This condition differs from the condition $k_1-k_2 = 2
\pi (\nu_{\rm hf}-\nu_{\rm t})/c$ of the undesired sideband
excitation $\ket{0g} \leftrightarrow \ket{1e}$ by the small amount
$4 \pi \nu_{\rm t}/c$, so the radiation linewidth must be tight
enough to prevent the undesired transition $\ket{0g}
\leftrightarrow \ket{1e}$. This condition is reached
experimentally by standard techniques which allow to reduce the
linewidths appropriately. Specific pulse shaping techniques can be
used to ensure an optimal probability of the correct transition
even for short pulses. Moreover, the logic states $\ket{0}$ and
$\ket{1}$ can be equivalently encoded in each of the two hyperfine
clock states. The swap scheme with optical transitions suits
better neutral $^{87}$Rb atoms trapped on atom chips than the
duplication scheme.

\section{Magnetic traps on atom chips and phase gate operation}
\label{magn}

To implement a phase gate between two qubit sites on an atom chip
we consider a very simple double well configuration.  The two
qubits are trapped in the two wells respectively, and they can
interact via the barrier.

\begin{figure}[t]
\begin{center}
\includegraphics[width=0.9 \columnwidth]{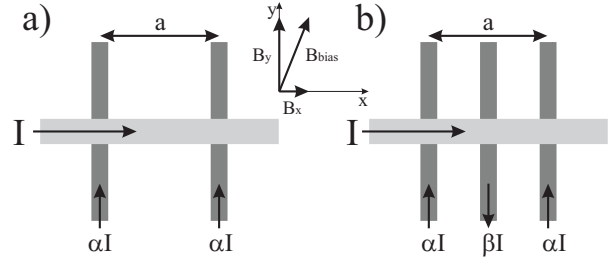}
\caption{\footnotesize{Wire configurations to create two qubit
traps on an atom chip, capable to achieve a gate operation. Tight
transverse confinement is created by the current I in the central
trapping wire and the transverse component of the bias field. The
qubit traps are then created by two crossing wires carrying a
smaller current $\alpha$I.  The angle of the bias magnetic field
is adjusted in such a way, as to achieve a magnetic offset fields
at the qubit trap locations equal to the magic field required for
trapped clock states (3.23G for Rb-87). {\em a)}: The H
configuration creates a double well potential to hold the qubits.
{\em b)}: Adding a third wire to the middle allows for flexible
control of the potential barrier height, and therefore more
control over the qubit coupling.}} \label{fig:geometry}
\end{center}
\end{figure}

In this paper we consider the simplest of these geometries for the
quantum gate implementation in magnetic micro traps on an atom
chip, the H configuration \cite{rei2} (see
Fig.\ref{fig:geometry}).  Here one wire along the $X$ axis ($X$
wire) carries a current $I$ and two parallel wires along the $Y$
axis (left and right wire), separated by a distance $a$, carry a
current $\alpha I$; a bias magnetic field ${\bf B_b}$ parallel to
the surface is employed. Appropriate values of currents and bias
field create the magnetic trapping  potentials for neutral atoms.
In the approximation of infinitely long and thin wires, with
negligible separation along the $Z$ axis, the $X$ wire and the
bias field produce a trapping field

\begin{equation}
{\bf{B}}^Q(x,y,z)=\left(B_b^x,B^y_b-\frac{\kappa\,I\,z}{z^2+y^2},\frac{\kappa\,I\,y}{z^2+y^2}\right),
\label{Bquadr}
\end{equation}
and the two parallel wires produce the localization fields

\begin{eqnarray}
{\bf{B}}^{\rm w}_L(x,y,z) & = & \frac{\kappa\,\alpha\,I}{z^2+(x-a/2)^2}\left(z,0,\frac{a}{2}-x\right)\nonumber\\
{\bf{B}}^{\rm w}_R(x,y,z) & = &
\frac{\kappa\,\alpha\,I}{z^2+(x+a/2)^2}\left(z,0,-\frac{a}{2}-x\right).
\label{BwLwR}
\end{eqnarray}
where $\kappa=\mu_0/(2\pi)$ and $\mu_0$ is the vacuum
permeability.

\begin{figure}[t]
\begin{center}
\includegraphics[width=8.0cm,height=5.0cm]{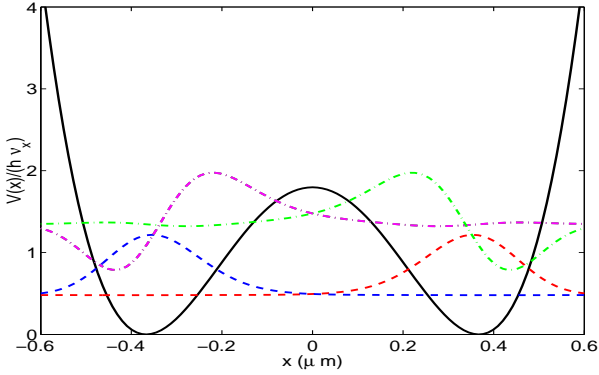}
\caption{\footnotesize{1D magnetic trapping potential for ${^{87}}$Rb
in the hyperfine states $\ket{F=1,m_F=-1}$,
$\ket{F=2,m_F=1}$. The distance is measured in $\mu$m; the trapping
potential is given in units
of $\hbar \omega_{x'}$ and shifted in order to vanish at the minima.
The ground (dashed) and first excited
(dash-dotted) vibrational levels in the left and right wells are shown.
Note that these states are linear combinations of the
eigenstates of the trapping potential.}}
\label{fig:potential}
\end{center}
\end{figure}

By using realistic parameters that have been tested in current
atom chips and setting $a=1.5$ $\mu$m, $I=29.9$ mA,
$\alpha=0.093$, $B^b_x=-9.91$ G, and $B^b_y=50$ G, we find a
trapping potential for two atoms (Fig.\ \ref{fig:potential}), with
two minima $(x_{0A},y_{0A})$ and $(x_{0B},y_{0B})$ lying on an
axis $X'$ rotated by an angle $\beta=\arctan (x_{0A}/y_{0A})
\simeq 0.063$ with respect to the $X$ axis. The magnetic field in
the two minima has the desired value $B_{M}=3.23$ G. The trap
frequencies in the two minima are $\omega_{x'}/(2\pi)\simeq 11.96$
kHz along the direction of longitudinal confinement,
$\omega_{y'}/(2\pi)\simeq 211.41$ kHz, and $\omega_z/(2\pi)\simeq
213.24$ kHz along the direction of transverse confinement. Atoms
in the transverse ground state will perform one-dim\-ension\-al
(1D) dynamics along the $X'$ axis only. The 1D potential is shown
in Fig.\ref{fig:potential}. The distance of the two minima from
the chip surface is $z_{0}\simeq 1.19$ $\mu$m; the separation of
the two minima is about $0.74$ $\mu$m. The variation of the
magnetic field along the trapping potential is small: at the
maximum height of the central barrier we have $B_{MAX}=3.26$ G.

We have used this static magnetic potential to study the
performance of a collisional phase gate. Steps (i) and (iii) of
the swap scheme require an optical two-photon transition, as
already discussed in Section \ref{exci}. The probability to
populate the second excited vibrational level $P_{20}=\mid
\bra{\varphi_2} e^{i\eta(a+a^\dagger)} \ket{\varphi_0} \mid^2$ is
about $3.8$ \% of the probability $P_{10}=\mid \bra{\varphi_1}
e^{i\eta(a+a^\dagger)} \ket{\varphi_0} \mid^2$ of sideband
excitation of the first excited vibrational level, thus affecting
the fidelity of the operation. In our simple scheme, the value
$3.8$ \% results from the anharmonicity of the longitudinal
trapping potential, and represents a lower limit that cannot be
reduced with a further decrease of the Lamb-Dicke parameter. In
the Lamb-Dicke regime the ratio of the two probabilities

\begin{equation}
P_{20}/P_{10} \simeq \frac{\mid \eta \bra{\varphi_2} (a+a^\dagger) \ket{\varphi_0} \mid^2}
{\mid \eta \bra{\varphi_1} (a+a^\dagger) \ket{\varphi_0} \mid^2}
\end{equation}
becomes independent of the Lamb-Dicke parameter. Nevertheless the
probability of this undesired transition can be further decreased
with an optimal modulation of the trapping potential in more
complex schemes that go beyond the simple gate investigated here.
For example, if one starts with a high barrier separating the two
wells, the vibrational states under considerations are much better
approximated by harmonic oscillator states $\ket{\varphi^{HO}_j}$
and thus the excitation of the higher vibrational states in the
Lamb-Dicke regime is essentially suppressed since
$\bra{\varphi^{HO}_j} a^\dagger \ket{\varphi^{HO}_0}
=\delta_{j1}$.

The realization of step (ii) of the swap scheme in the trapping
potential in Fig.\ref{fig:potential} can also be considered as a
simplified version of the phase gate proposed in \cite{will}. The
ground and first excited vibrational states of each potential well
are the logic states. These states are linear superpositions of
eigenstates of the trapping potential; free atoms would tunnel
from one well to the other. The presence of one atom in each well
and their mutual interaction when they collide modifies
significantly the dynamics. The choice of the trapping potential
of Fig.\ref{fig:potential} combines two opposite requirements: the
ground states must be deep in the well to avoid tunneling during
the gate operation, while the excited states have to interact in
order to develop the correct value of the phase $\phi$.

\begin{figure}[t]
\begin{center}
\includegraphics[width=8.0cm,height=5.0cm]{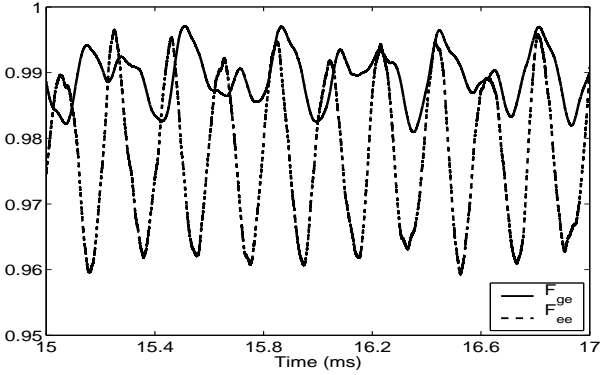}
\caption{\footnotesize{Revivals of the vibrational states $\ket{\Psi_{ge}}$ (solid line) and
$\ket{\Psi_{ee}}$ (dashed line) in a static magnetic double well potential during the phase gate operation.}}
\label{fig:overlap}
\end{center}
\end{figure}

\begin{figure}[t]
\begin{center}
\includegraphics[width=8.0cm,height=5.0cm]{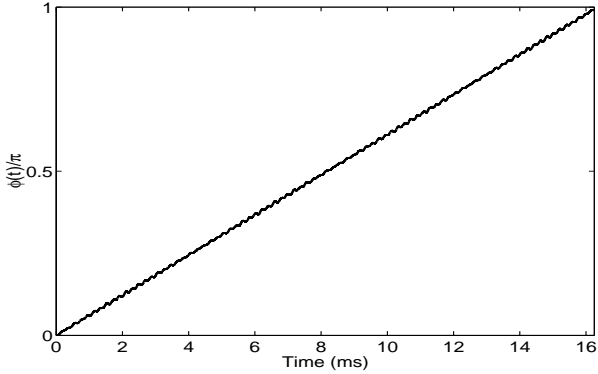}
\caption{\footnotesize{Dynamics of the phase $\phi$ during the gate operation.}}
\label{fig:phase}
\end{center}
\end{figure}

We have already discussed how to transfer the logic states from
the storage qubit to the operation qubit. We examine now the gate
operation performed with the vibrational states.
Initially one atom sits in each well and the phase gate operation
is performed via tunneling of the first excited states of the two
atoms. A correct operation of the phase gate requires a complete
revival (up to a phase) of each of the initial states
$\ket{\Psi_{gg}}$, $\ket{\Psi_{ge}}$, $\ket{\Psi_{eg}}$ (this
state is simmetric of $\ket{\Psi_{ge}}$ so we shall neglect it)
and $\ket{\Psi_{ee}}$ at some later time $\tau$, as well as the
fulfillment of the condition
$\phi=\phi_{ee}+\phi_{gg}-2\phi_{ge}=\pi$ ($\phi_{ab}$ is the
phase of the state $\ket{\Psi_{ab}}$) at the same time $\tau$. We
solve the two-particle Schr\"odinger equation numerically in 1D with the split
operator technique \cite{spli}, replacing the scattering length
$a_s$ of the hyperfine state $\ket{F=2,m_F=1}$ of $^{87}$Rb with
an effective scattering length $a_\perp$ that takes into account
the transverse confinement \cite{cala}. The interaction between the
atoms is described by a contact potential, like in \cite{cala,tren}. After an operation time
$\tau \simeq 16.25$ ms, the initial states $\ket{\Psi_{ge}(0)}$
and $\ket{\Psi_{ee}(0)}$ have an almost complete revival, with
fidelities $F_{ge}=\mid \langle \Psi_{ge}(0) \mid \Psi_{ge}(\tau)
\rangle \mid^2 > 0.99$, $F_{ee}=\mid \langle \Psi_{ee}(0) \mid
\Psi_{ee}(\tau) \rangle \mid^2 > 0.99$ (see
Fig.\ref{fig:overlap}); moreover, the state $\ket{\Psi_{gg}}$ is
stationary on this time scale.
We have also evaluated the gate phase accumulated during the gate operation, and the final result
is $\phi \simeq 0.99 \pi$ (see Fig.\ref{fig:phase}).

\begin{figure}[t]
\begin{center}
\includegraphics[width=8.0cm,height=5.0cm]{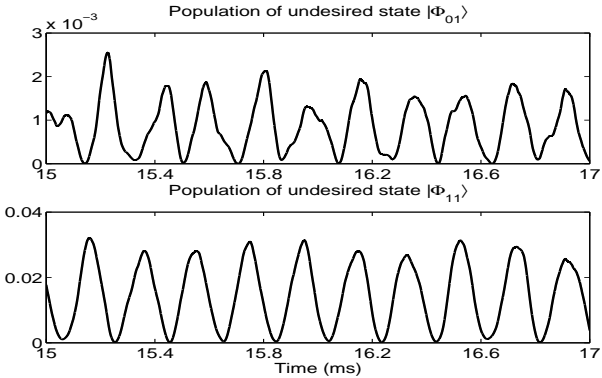}
\caption{\footnotesize{Populations of vibrational states $\ket{\Phi_{ge}}$ (upper figure) and
$\ket{\Phi_{ee}}$ (lower figure) of atoms sitting in the same potential well.}}
\label{fig:undesired}
\end{center}
\end{figure}

The populations of undesired states $\ket{\Phi_{ge}}$ and $\ket{\Phi_{ee}}$, where the atoms are in the
ground and first excited vibrational level of the same well, must be negligible.
Our calculations confirm that these populations remain small (see Fig.\ref{fig:undesired}). These
results also show that the effects of tunneling, that would increase the populations of the undesired
states in the absence of interaction, are significantly reduced, as already stressed in \cite{will}.

These results are encouraging, since they are obtained with
realistic parameters and without introducing approximations for
numerical convenience. The use of the clock states
$\ket{F=1,m_F=-1}$ and $\ket{F=2,m_F=1}$ greatly reduces the
impact of fluctuations of the magnetic field due to current
fluctuations, that would spoil the gate performance. Moreover, the
use of a static trapping potential has the advantage of avoiding
the necessity of accurate control over the relevant parameters
(currents, magnetic fields, etc.) as requested by dynamic schemes.
More complex approaches might further increase gate performances
and reduce the operation time. We briefly discuss them in the next
section.

\section{Future perspectives}
\label{futu}

Atom chips provide the setup for a rich variety of traps for
neutral atoms. In this paper we have discussed the simplest static
magnetic potential for implementing quantum gates. Much more
involved wire configurations can be thought of, electric fields
coupling to the electric polarizability of the atom
($U_{el}=-\frac{1}{2} E^2$) can be brought in as additional design
freedom \cite{last}, RF and MW fields coupling different atomic
states can be used to create adiabatic potentials and (slowly
varying) time-dependent potentials will increase the versatility
of qubit manipulation.

A first extension will be to use a three wire configuration as
shown in Figure \ref{fig:geometry}b which allows more freedom in
designing the double well potential for qubit coupling. The wire
in the middle can be used to control the barrier either by current
or by electric fields. A time varying barrier potential height
will allow techniques from quantum control \cite{Skl02} to be
applied, and we expect a significant increase in the speed and
fidelity of the quantum gate.

Encoding the qubit in different hyperfine states, like
$\ket{F=2,m_F=1}$ and $\ket{F=2,m_F=2}$ of $^{87}$Rb, which have
different interactions with magnetic fields but which couple
equally to electric fields, will allow to use electric fields for
state dependent collisions in two qubit operations. A combination
of both magnetic and electric interactions allows to have a
barrier between the $\ket{F=2,m_F=2}$ state, and a collisional
interaction between the $\ket{F=2,m_F=1}$ states, thus opening the
possibility of state-depend\-ent dynamics. The drawback here is
that we now have to deal with field sensitive qubits. This implies
a demand for excellent magnetic field stability.

\begin{figure}[t]
\begin{center}
\includegraphics[width=\columnwidth]{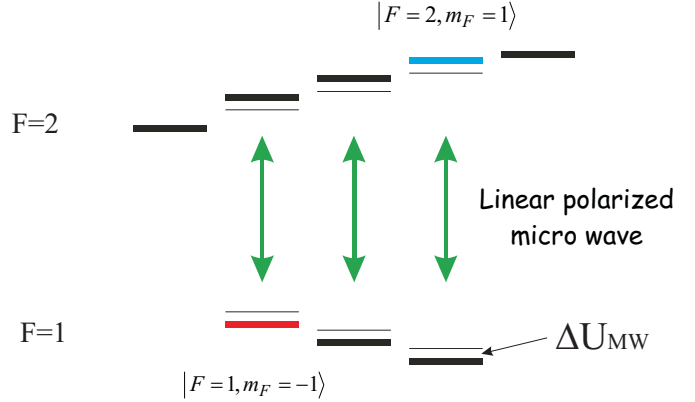}
\caption{\footnotesize{Creation of qubit state selective micro
wave potentials.  Applying a linar polarized MW detuned from the
hyperfine transition creates opposite potential shifts for the two
hyperfine manifolds.  The two trappable qubit states
$\ket{F=2,m_F=1}$ and $\ket{F=1,m_F=-}$ of $^{87}$Rb are
indicated.}} \label{fig:MWpot}
\end{center}
\end{figure}

Even more versatile manipulation can be achieved by using the
adiabatic potentials created by RF and MW fields to manipulate
atoms \cite{garra}. In an atom chip setup such RF or MW potentials
have the distinct advantage, that one applies RF and MW near
fields with a specific well defined polarization which is given by
the wire configuration on the chip. At the relevant scales of the
atom chip ($\mu m$) propagation effects can be neglected and the
fields can be calculated in a quasi static approximation.

The coupling strength to the MW (RF) field is then given by the
Rabi frequency $\hbar \Omega_{MW} = \overrightarrow{\mu} \cdot
\overrightarrow{B}_{MW}$.  In the case of large detuning
($\Delta_{MW} \gg \Omega_{MW}$) the adiabatic potential can then
be calculated as $U_{MW}= \pm \hbar \Omega^2 _{MW} / 4
\Delta_{MW}$. Applying linear polarized MW radiation detuned from
the hyperfine transition will create opposite potential shifts for
the two hyperfine manifolds. This result in opposite sign MW
potentials for two clock states $\ket{F=2,m_F=1}$ and
$\ket{F=1,m_F=-1}$ of $^{87}$Rb \cite{rei1}. Applying this MW field to a wire
between the two trapping wires (the center wire in the 3 wire
configuration of Fig.~\ref{fig:geometry}b) it will be possible to
create state dependent potential barriers on the chip, which will
greatly enhance the qubit manipulation capabilities.

\begin{figure}[t]
\begin{center}
\includegraphics[width=\columnwidth]{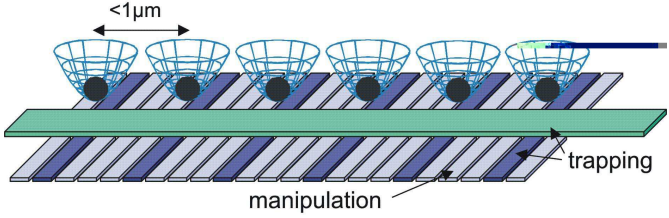}
\caption{\footnotesize{A schematic drawing od an array atom chip
quantum processor.  Atoms are trapped with strong transverse
confinement along a current carrying wire.  The qubit locations
are determined by transverse crossing wires.  In between are
manipulation wires which allow to control the barriers between the
qubit sites.}} \label{fig:ACQP}
\end{center}
\end{figure}

The above simple two qubit potentials and operations can be
geralized to a atom chip quantum processor on an array of qubits.
We envision such more advanced qubit manipulation on the atom chip
(Fig.~\ref{fig:ACQP}) to consist of an array of trappable qubits,
consisting of single atoms trapped in tight traps. The strong
transverse confinement will be created by a tight magnetic wire
trap, the qubit sites can then be determined either by electrical
leads or by crossed wires, as in the case considered here. In
between these trapping sites we will have additional wires and
electrodes for qubit manipulation and gate operations. these would
create the barriers for qubit coupling and qubit isolation.  At
the end the qubit array can be read out by transporting the atoms
one by one to an integrated micro cavity atom detector
\cite{horak03}, reading their internal state.

\section{Summary and Conclusions}
\label{conc}

We have discussed new simple scheme for implementing a phase gate
on an atom chip. These schemes make use of two degrees of freedom
for each qubit. Quantum information is stored in one degree of
freedom and processed in the other. We have discussed the
feasibility of realizing these schemes with cold $^{87}$Rb atoms
magnetically trapped on atom chips, where the internal states are
used to storage and the external (vibrational) states to process
information, respectively. We have also investigated the
performance of such schemes in atom chip traps with realistic
parameter values. High fidelity gate operations are obtainable,
thus showing that micro traps on atom chips are a very interesting
candidate for quantum engineering and quantum information
processing. Our investigations have concentrated on the most basic
example of the many potential applications of micro traps on atom
chips. Some more involved schemes are discussed as possible future
developments.

\begin{acknowledgement}
This work was supported by the European Union, contract numbers
IST-2001-38863 (ACQP), MRTN-CT-2001-50532 (AtomChips), and the
Deutsche Forsch\-ungsgemeinschaft, Schwerpunktprogramm
`Quanteninformations\-verarbeitung'. We thank M. Andersson, E.
Charron, A. Recati, P. Treutlein for useful discussions.

\end{acknowledgement}


\begin{thebibliography}{99}
\bibitem{comm} The first term of the series expansion $e^{ikx}=1+ikx+\ldots$ leads to a new
definition of the internal eigenstates and their frequencies. For ease of notation, we continue
to denote them with $\ket{i}$, $\ket{f}$ and $\omega_i$, $\omega_f$.
\end{thebibliography}
\end{document}